\documentclass[twocolumn,amsmath,amssymb,pra,superscriptaddress]{revtex4}

\usepackage{graphics}
\usepackage{epsfig}
\usepackage{amsmath}
\usepackage{amsfonts}
\usepackage{color}
\usepackage{dcolumn}
\usepackage{bm}
\usepackage{mathrsfs}

\begin{document}

  \bibliographystyle{apsrev}
  
\title{Partitioning of the Electron Correlation Energy}

      \author{Eric Ouma Jobunga}
       \affiliation{\it  Department  of Mathematics and Physics , Technical University   of Mombasa,\\ P. O. Box 90420-80100, Mombasa, Kenya}
%
%


\begin{abstract}
Electron-electron correlation forms the basis of difficulties encountered in many-body problems. Accurate treatment of the correlation problem is likely to unravel some nice physical properties of matter embedded in this correlation. In an effort to tackle this many-body problem, an exact partition function for the correlation energy between two interacting states is proposed in this study. Using this partition function, a model potential for a two-electron system is derived. The model potential can accurately reproduce the singlet states of a singly excited neutral helium atom with one of its electrons frozen in the $1s$ orbital.

\end{abstract}

\maketitle
\section{Introduction}

Helium atom and helium-like ions are the simplest many-body systems containing two electrons which interact among themselves in addition to their interaction with the nucleus. The two-electron systems are therefore the ideal candidates for studying the  electron correlation effects. 

 Several theoretical approaches have been employed in the past in dealing with the electron correlation problem. Some of the approaches include the  variational Hyleraas method \cite{Hylleraas1929, Drake1999}, coupled channels method \cite{Barna2003}, the configuration interaction method \cite{Hasbani2000}, explicitly correlated basis and complex scaling method \cite{Scrinzi1998}. At present only the Hylleraas method, which includes the interelectronic distance as an additional free co-ordinate, yields the known absolute accuracy of the groundstate energy of the helium atom \cite{Pekeris1959}. 
 
 Configuration interaction methods have also been proved to be accurate but they are quite expensive computationally. To overcome the computational challenges especially for really large systems, single active electron (SAE) methods become advantageous, although some approximations are necessary in developing the model potentials \cite{Parker2000, Parker1998}. Reasonably accurate eigenvectors and energies can be generated using the model potentials. 
 
  The development of the SAE potentials is an active field of study taking different approximations \cite{Chang1976} like the independent particle approximation (IPA), multi-configurational Hartree-Fock (HF) \cite{Szabo1996}, density functional theory (DFT) \cite{Kohn1965}, random phase approximation (RPA) \cite{Linderberg1980}, and many others . The major limitation of SAE approximations is the inability to explain multiple electron features like double excitation, simultaneous excitation and ionization, double ionization, and innershell transitions. However, progress is being made towards the realization of these features.

The non-relativistic Hamiltonian of a two-electron system with a nuclear charge $Z$ is given by

    \begin{equation}
    \mathrm{H} = \frac{1}{2}\, \left[p_1^2 + p_2^2\right] - Z\, \left[\frac{1}{r_1} + \frac{1}{r_2} \right] + \frac{1}{|\mathbf{r}_1-\mathbf{r}_2|}
    \end{equation}
where the first term correspond to the sum of the kinetic energy of each of the two electrons, the second term to the sum of the interactions between each of the electrons and the nucleus, and the last term to the electron correlation interaction between the two electrons. The second and the last term form the potential energy function of a bound two-electron system.

 In our previous work \cite{Jobunga2017}, it was shown that the electron correlation interaction analytically simplifies to
 \begin{equation}
 \frac{1}{|\mathbf{r}_i-\mathbf{r}_j|} = \frac{1}{\sqrt{r_i^2 + r_j^2}}
 \end{equation}
 because of orthogonality of the two interacting quantum states. In the independent particle approximation method, the potential function 
 \begin{equation}
 V(r_i, r_j) = -\frac{Z}{r_i} + \frac{1}{2}\,\frac{1}{\sqrt{r_i^2 + r_j^2}} \label{eq:pt1}
 \end{equation}
for a two-electron system, using the mean field, can be completely separated \cite{Jobunga2017} as
\begin{equation}
 V(r_i) = -\frac{Z}{r_i} + \frac{1}{2}\,\frac{\sqrt[3]{2Z}}{r_i} \label{eq:pt1b}.
 \end{equation}
 Factor $1/2$ in equation (\ref{eq:pt1}) assumes an equal sharing of the correlation energy between the two interacting quantum states. While this may be true if the interacting states are identical or degenerate, the validity of the equal sharing assumption is lost for a pair non-degenerate interacting states. In ref.\cite{Jobunga2017}, it can be seen that accurate groundstate energy of helium atom and reasonable eigenvalues of autoionizing levels of identical symmetry have been obtained using the equal partitioning of the correlation energy. On the other hand, singly excited states of helium atom with one electron frozen in the ground state is poorly represented by this partioning of the correlation energy. 
 
 In this work, a more appropriate energy sharing relation based on the geometry of the problem is proposed. This is consequently used to approximate the average correlation energy per eigenstate in terms of separable electron co-ordinates. To test the accuracy of the argument, a SAE model potential for two-electron systems is suggested and optimized for a helium atom whose eigenvalues are well known.  The method advanced is quite original and we do not have any knowledge of a similar work elsewhere.
 
\section{Theory} 
  From the physics of oscillations, it is known that the potential energy of a vibrating particle is proportional to the square of the amplitude of vibration.  That is, ${\epsilon = 1/2\,k\,r^2}$, where $k$ in this case is equivalent to a spring constant, and $r$ is the vibration amplitude. The total potential energy for two interacting electrons would therefore be given by ${\epsilon_{tot}=1/2\,k\,[ r_i^2 + r_j^2] }$. It can be hypothesized that the correlation energy between two interacting electrons is shared in proportion to their corresponding potential energies. That is, the correlation energy due to the $i^{\mathrm{th}}$ electron is,
 \begin{equation}
 \begin {split}
 \delta E_i^{\mathrm{corr}} &= \frac{r_i^2}{r_i^2 + r_j^2}\, \delta E_{\mathrm{tot}}^{\mathrm{corr}} \\
                            &= \frac{r_i^2}{r_i^2 + r_j^2}\, \frac{1}{\sqrt{r_1^2 + r_2^2}} \label{eq:pf}
 \end{split}                           
 \end{equation}
 where $ \delta E_{\mathrm{tot}}$ is the total correlation energy as given in equation (\ref{eq:pt1b}) and its prefactor is the partition function.
 The potential function describing the $i^{\mathrm{th}}$ electron in the independent particle approximation can then be expressed as
 \begin{equation}
 V(r_i, r_j) = -\frac{Z}{r_i} + \frac{r_i^2}{\left\lbrace r_i^2 + r_j^2 \right\rbrace^{\frac{3}{2}}}. \label{eq:pt2}
 \end{equation}
Equation (\ref{eq:pt2}) if minimised with respect to $r_i$ leads to
\begin{equation}
 \frac{\partial V(r_i, r_j)}{\partial r_i} = \frac{Z}{r_i^2} + \frac{2\,r_i}{\left\lbrace r_i^2 + r_j^2 \right\rbrace^{\frac{3}{2}}} - \frac{3\,r_i^3}{\left\lbrace r_i^2 + r_j^2 \right\rbrace^{\frac{5}{2}}} = 0. \label{eq:pt3}
 \end{equation}
 or 
\begin{equation}
  \frac{Z}{r_i^2}  + \frac{2\,r_i\,r_j^2-r_i^3}{\left\lbrace r_i^2 + r_j^2 \right\rbrace^{\frac{5}{2}}} = 0, \label{eq:pt4}
 \end{equation}
as the condition for an extremum potential. Equation (\ref{eq:pt4}) can be reorganized further by reversing the sign of the coefficient of $r_i\,r_j^2$ and incrementing the coefficient of $r_i^3$ by $1$. The reorganization introduces an inequality
 \begin{equation}
  \frac{Z}{r_i^2}  - \frac{2\,r_i\,r_j^2+ 2\,r_i^3}{\left\lbrace r_i^2 + r_j^2 \right\rbrace^{\frac{5}{2}}} \leq 0, \label{eq:pt5}
 \end{equation}
  which ensures the potential is minimized while treating the co-ordinates $r_i$ and $r_j$ with an equal weighting. The equality condition in equation (\ref{eq:pt5}) guarantees a minimum potential. It is from this condition that the correlation term
 \begin{equation}
 \frac{1}{\sqrt{r_i^2 + r_j^2}} = \frac{1}{r_i}\left[\frac{Z}{2}\, \frac{r_i^2}{r_i^2 + r_j^2} \right]^{\frac{1}{5}}
 \end{equation}
 is evaluated and equation (\ref{eq:pt2}) simplifies as
 \begin{equation}
 V(r_i, r_j) = -\frac{Z}{r_i} + \frac{\left[(Z/2)\,f(r_i,r_j)\right]^{\frac{3}{5}}}{r_i} \label{eq:pt6}
 \end{equation}
 where the correlated two-dimensional function 
 \begin{equation}
 f(r_i, r_j) = \frac{r_i^2}{r_i^2 + r_j^2}
 \end{equation}
 is equivalent to the partition function already introduced in equation (\ref{eq:pf}). With regards to SAE, the value of the function ${f(r_i,r_j)^{\frac{3}{5}}}$ in equation (\ref{eq:pt6}) cannot be evaluated exactly but can only be approximated by taking its expectation value relative to the some trial wavefunction of the $j^{\mathrm{th}}$ electron. In our case, we have used the hydrogenic wavefunction of the $1s$ orbital as the trial wavefuction and the conditions ${0 \leq r_j \leq r_i }$ and ${r_i \leq r_j \leq \infty}$ in evaluating the expectation value of the function in terms of the radial co-ordinate $r_i$.
 
In our working, the expectation value of the correlated function, expressed in terms of one of the radial co-ordinate, is evaluated approximately as
\begin{equation}
 \langle f(r_i, r_j)^{\frac{3}{5}} \rangle \approx 1 - \left[ \frac{27}{25} + \frac{6}{5}Zr_i - \frac{6}{125\,Zr_i}\right]\, \exp (-2Zr_i). \label{eq:pt7}
 \end{equation}
  Appendix A  shows the explicit method used in arriving at this expectation value.
 A further empirical and intuitive optimization of the expectation value given by equation (\ref{eq:pt7}) is employed to obtain 
 \begin{equation}
 \langle f(r_i, r_j)^{\frac{3}{5}} \rangle \approx 1 - \alpha\,\left[ 1 + 3Zr_i\right]\, \exp (-2Zr_i) \label{eq:pt8}
 \end{equation}
 with the parameter ${\alpha=0.46135}$ set to include other significant corrections. The approximation in equation (\ref{eq:pt8}) if employed in the independent electron potential, defined in equation (\ref{eq:pt6}), is found to be of a better agreement with the experimental results for the singly excited helium atom as compared to equation (\ref{eq:pt7}). Substituting the expectation value obtained into equation (\ref{eq:pt6}) reduces the correlated problem into a single electron model potential
 \begin{equation}
 V(r_i) = -\frac{Z}{r_i} + \frac{\left[Z/2\right]^{\frac{3}{5}}\,\zeta(r_i)}{r_i} \label{eq:pt9}
 \end{equation}

 with $\zeta(r_i)$ given by equations (\ref{eq:pt7}) or (\ref{eq:pt8}).  With this potential, the SAE Hamiltonian 
 \begin{equation}
 H(r_i)= \frac{p_i^2}{2} + V(r_i) \label{eq:pt10}
 \end{equation}
 
 is defined. It is evident that the first term of the SAE potential defined in equation (\ref{eq:pt9}) is the electron-nuclear interaction, and the second term yields the screening potential of the active electron from the other electron.  The eigenvalues of a two-electron system can then be evaluated as \cite{Jobunga2017}
 \begin{equation}
  \langle E_{\alpha \alpha'} \rangle  = \left\{ \begin{matrix}
  4\, {\varepsilon}_{\alpha \alpha'} & \mathrm{if}\; \alpha = \alpha'\\
  {\varepsilon}_{\alpha \alpha} + {\varepsilon}_{\alpha' \alpha'} & \mathrm{if}\; \alpha \neq \alpha'
  \end{matrix} \right.
 \end{equation}
 where ${{\varepsilon}_{\alpha \alpha} = \langle H(r_i) \rangle }$ is the eigenvalue of a single electron orbital.  Factor $4$ in the above equation arises from both exchange and degeneracy consideration for states with $\alpha=\alpha'$. For a helium atom with one electron considered to be in the ground state and the other electron occupying an excited state $\alpha'$, ${\varepsilon}_{\alpha \alpha}$  is approximately equal to the core energy eigenvalue, ${E_{\mathrm{core}}=-2.00000}$, for the helium ion in its ground state.
 
\section{Results and Discussions}

We have developed a single active electron (SAE) model potential for a two-electron system given by equation (\ref{eq:pt9}). The model potential is used to calculate the $1snl$ eigenvalues for helium atom as shown in table \ref{tab1} for angular momenta of up to ${l_{\mathrm{max}}=7}$. In the table, we use the two alternative relations expressed in equations (\ref{eq:pt7}) and (\ref{eq:pt8}) to evaluate the eigenvalues given by $H_1$ and $H_2$ respectively. The results are presented for the first five principal quantum numbers for each angular momentum values. In generating our results, a B-spline radial box of $600$ B-splines, ${r_{\mathrm{max}}=200}$, $k=10$, and a non-linear knot sequence is used.

\begin{table}[!ht]
    \centering
    \begin{tabular}{cccc}
    \hline
    State & $H_{1}$ & $H_{2}$   & Ref.  \\
   \hline
   \hline
       $L=0$&-3.29443   & -2.90367 & -2.90372   \\                                                     
            &-2.15290   & -2.14580 & -2.14597   \\                          
            &-2.06325   & -2.06136 & -2.06127  \\  
            &-2.03439   & -2.03364 & -2.03358   \\ 
            &-2.02158   & -2.02120 &    \\                                  
    \hline
       $L=1$&-2.12631   & -2.12617 &  -2.12384  \\                                                     
            &-2.05600   & -2.05595 &  -2.05514  \\                          
            &-2.03144   & -2.03142 &  -2.03106  \\  
            &-2.02010   & -2.02009 &  -2.01991  \\ 
            &-2.01394   & -2.01394 &    \\                                  
    \hline
       $L=2$&-2.05555   & -2.05555 &  -2.05562  \\                                                     
            &-2.03125   & -2.03125 &  -2.03127  \\                          
            &-2.02000   & -2.02000 &  -2.02001  \\  
            &-2.01388   & -2.01388 &  -2.01389  \\ 
            &-2.01020   & -2.01020 &    \\                                  
    \hline
       $L=3$&-2.03125   & -2.03125 &   -2.03125 \\                                                     
            &-2.02000   & -2.02000 &   -2.02000 \\                          
            &-2.01388   & -2.01388 &   -2.01389 \\  
            &-2.01020   & -2.01020 &   -2.01020 \\ 
            &-2.00781   & -2.00781 &    \\                                  
    \hline
       $L=4$&-2.02000   & -2.02000 &   -2.02000 \\                                                     
            &-2.01388   & -2.01388 &   -2.01388\\                          
            &-2.01020   & -2.01020 &   -2.01020 \\  
            &-2.00781   & -2.00781 &    \\ 
            &-2.00617   & -2.00617 &    \\                                  
    \hline
       $L=5$&-2.01388   & -2.01388 &   -2.01388 \\                                                     
            &-2.01020   & -2.01020 &   -2.01020 \\                          
            &-2.00781   & -2.00781 &   -2.00781 \\  
            &-2.00617   & -2.00617 &    \\ 
            &-2.00500   & -2.00499 &    \\                                  
    \hline
       $L=6$&-2.01020   & -2.01020 &   -2.01020 \\                                                     
            &-2.00781   & -2.00781 &   -2.00781 \\                          
            &-2.00617   & -2.00617 &   -2.00617 \\  
            &-2.00499   & -2.00500 &    \\ 
            &-2.00413   & -2.00413 &    \\                                  
    \hline
       $L=7$&-2.00781   & -2.00781 &   -2.00781 \\                                                     
            &-2.00617   & -2.00617 &   -2.00617 \\                          
            &-2.00499   & -2.00500 &   -2.00499 \\  
            &-2.00413   & -2.00413 &    \\ 
            &-2.00347   & -2.00347 &    \\                                  
    \hline
    \end{tabular}
    \caption{Some numerically calculated eigenvalues using the present model potentials versus the reference values for helium atom \cite{Scrinzi1998}.  The $H_{1}$ and $H_2$ are the SAE Hamiltonian with equations (\ref{eq:pt7}) and (\ref{eq:pt8}) as the model potentials respectively. The results presented are truncated at $6$ s.f. }
    \label{tab1}
  \end{table}

The results generated with the model potentials presented are in good agreement with the references values \cite{Scrinzi1998} at larger values of $n$ and $l$ as expected.  At these higher quantum numbers, the spatial extent of the orbitals is larger reducing the significance of the electron-electron interaction. In particular, one can see that for ${l \geq 2}$, the three sets of results are in good agreement with each other. The discrepancy between the set of results essentially manifest at lower values of $n$ and $l$. The disparities are quite evident for ${l=0}$ and ${l=1}$ states presented. These lower angular momentum states usually provide the stringest test of accuracy for any model potential. 

As can be seen, the groundstate yields the largest deviation in the results. The $H_1$ interaction yields an unphysical tight binding potential to the groundstate helium atom. This emanates from the confinement introduced by the shortrange term $1/r^2$ in the correlation term. In the $H_2$ interaction, the shortrange confinement is removed. The eigenvalues generated using the model potential in $H_2$ are in good agreement with reference values.

The removal of the shortrange interaction in the SAE model potential in $H_2$ is motivated by its absence in the exactly separable symmetric term $V_{\alpha \alpha}$ as given in equation (\ref{eq:pt1b}). The good agreement between the eigenvalues in the model potential of equation (\ref{eq:pt8}) and the reference results attests to the credibility of the method introduced in ref. \cite{Jobunga2017} and advanced in this paper.

\section{Conclusion}
The exact partitioning of the correlation energy between two interacting electrons in different orbitals is tackled in this paper. A partition function which depends on the spatial extent of the interacting states is suggested. In our working, we obtain a non-separable correlated function whose expectation value is approximated as a function of one of the radial orbitals. This leads to separability of the corrrelated term in the two-electron Hamiltonian. The partition function introduced and the consequently optimized SAE model potential developed in this work proves to be reasonably accurate in the calculated eigenvalues. The proposed model potential can be extended further to other helium-like systems by adjusting the $\alpha$ parameter to reproduce the experimentally known groundstate potential of the respective system.

\appendix
\section{}

The method through which the expectation value in equation (\ref{eq:pt7}) has been evaluated is shown in this appendix. The integral 
\begin{equation}
\begin{split}
 \langle f(r_i, r_j)^{\frac{3}{5}} \rangle =& \langle \phi (r_j)|\left[\frac{r_j^2}{r_i^2 +r_j^2}\right]^{\frac{3}{5}}|\phi(r_j)\rangle\\
           =& \int_{0}^{r_i} \mathrm{d}r_j\, \left[r_j^2\, t^{\frac{6}{5}}\left(1 + t^2\right)^{-\frac{3}{5}}\right]\exp (-2Zr_j)\\ & + \int_{r_i}^{\infty} \mathrm{d}r_j\,\left[r_j^2\, \left(1 + t^2\right)^{-\frac{3}{5}}\right]\exp(-2Zr_j) \label{eq:app1}
 \end{split}
 \end{equation}
 is evaluated in parts where we consider that ${0\leq r_j \leq r_i}$, ${r_i \leq r_j \leq \infty}$, ${t=r_</r_>}$, ${r_<=\mathrm{min}(r_i, r_j)}$, and ${r_>=\mathrm{max}(r_i, r_j)}$. We have used the hydrogenic orbital ${\phi(r_j) = \exp(-Zr_i)}$ and a binomial expansion of
\begin{equation}
(1 + t^2)^{-\frac{3}{5}} = \sum_{k=0}^{\infty} \left(\begin{matrix} {-3/5}\\ k \end{matrix} \right)\, t^{2k}   \label{eq:app2}
\end{equation} 
  to evaluate this expectation value assuming that one of the electrons is localized inside the ground state ionic core. Equation (\ref{eq:app1}) together with the series in equation (\ref{eq:app2}) yield an integral that cannot be evaluated exactly. In our case, only $k=0$ and $k=1$ are used for estimation. It is important to note that the expectation value in this case provides a static contribution of the correlated term to the active electron in the field of the other electron.
  \\ 
  \\


\bibliographystyle{apsrev}
\bibliography{/home/eric/Inworks/Literature}

\end{document}